\documentstyle[twoside,fleqn,espcrc2]{article}

\newcommand{\half}{{{1\over  2} }}
\newcommand{\CZ}{{\cal{Z}}}
\newcommand{\CH}{{\cal{H}}}
\newcommand{\CR}{{\cal{R}}}
\newcommand{\CS}{{\cal{S}}}
\newcommand{\CF}{{\cal{F}}}

\newcommand{\CL}{{\cal{L}}}
\newcommand{\CM}{{\cal{M}}}
\newcommand{\CO}{{\cal{O}}}
\newcommand{\CP}{{\cal{P}}}
\newcommand{\mod}{{\rm mod}}
\newcommand{\Tr}{{\rm Tr}}
\newcommand{\lieg}{{\underline{\bf g}}}
\newcommand{\liet}{{\underline{\bf t}}}

\newcommand{\AmS}{{\protect\the\textfont2
  A\kern-.1667em\lower.5ex\hbox{M}\kern-.125emS}}

\title{String duality, automorphic forms, and generalized
Kac-Moody algebras}

\author{Gregory Moore \address{Department of Physics,
        Yale University, \\
        Box 208120, New Haven, CT 06520, USA}%
        \thanks{Supported by
DOE grant DE-FG02-92ER40704.}
}

\begin{document}

\begin{abstract}
We review some recent work that has been done
on the relation of BPS states, automorphic forms
and the geometry of the quantum ground states of
string compactifications with extended supersymmetry.
\end{abstract}

\maketitle

\section{Introduction }

In the past two years it has become
clear that there are some interesting relations between
string duality, the theory of generalized Kac-Moody
algebras (GKM's),  and their associated automorphic
forms. These connections are of three kinds:

\smallskip
\noindent
1. The space of BPS states is an algebra, sometimes
it is a GKM.

\smallskip
\noindent
2. Black hole BPS states are counted by a
denominator product for a GKM

\smallskip
\noindent
3. Threshhold corrections involve automorphic
forms similar to those associated to GKM's.
\smallskip

Unfortunately, the precise relations between these three
occurances are still shrouded in mystery. This talk
reviewed these three circles of ideas, emphasizing
a point of view we developed in a series of papers with
J. Harvey \cite{hmi,hmii,hm96b,hm96c}.

\section{ The algebra of BPS states  }

In theories of extended supersymmetry the
space of BPS states canonically forms an
algebra \cite{hmi,hmii}.

\subsection{Definition}

Let us recall the definition from \cite{hmi,hmii}.
We assume that there are exactly conserved
 charges $Q$ so that

a.) The Hilbert space splits as a sum of  superselection sectors: $\CH =
\oplus_{Q} \CH^Q$

b.) In each sector there is a
  Bogomolnyi bound: $E\geq \parallel \CZ(Q)\parallel $.
Here $\parallel \CZ(Q)\parallel$ is the invariant norm of
the central charge in $ \CH^Q$.

In each sector there is a  distinguished subspace
of ``BPS states'':
\begin{equation}
\CH_{BPS} \equiv \oplus_{Q} \CH_{BPS}^Q
\end{equation}
defined by the space of states saturating the bound. Note that
with this definition the space of BPS states includes
multi-particle (noninteracting) states in addition to one-particle
boundstates at threshhold.

The algebra structure
\begin{equation}
  \CR: \CH_{BPS} \otimes \CH_{BPS} \rightarrow \CH_{BPS}
\end{equation}
is defined as follows. We consider the $S$-matrix
$\CS(\psi_1 + \psi_2 \rightarrow \CF)$ where $\CF$ is
an arbitrary final state. Under analytic continuation of
the total center of mass-squared the $S$-matrix
has a distinguished pole:
\footnote{our signature is $(-,+^{d-1})$ in $d$ spacetime
dimensions, and $s$ is the usual Mandelstam variable.}
\begin{equation}
\CS(\psi_1 + \psi_2 \rightarrow \CF)
\sim {Res
\over
s + \parallel \CZ(Q_1+Q_2)\parallel^2}
\end{equation}
 Since
\begin{equation}
 E(\vec p) \geq \parallel {\CZ}(Q_{1})\parallel +
\parallel \CZ(Q_2) \parallel \geq \parallel \CZ(Q_1 + Q_2) \parallel
\end{equation}
finding the  pole requires analytic continuation unless there are
boundstates at threshhold in the sector $Q_1 + Q_2$.
The residue of the pole
must be expressed in terms of matrix elements with
{\it on-shell} states. By charge conservation and
the Bogomolnyi bound these states must be the
BPS states in the superselection sector
$Q_1 + Q_2$. Therefore the residue of the
pole ``factors through'' the space of BPS states:
\begin{equation}
Res =\sum_{\psi_3^I \in \CH_{BPS}^{Q_1+Q_2}  }
\langle \CF \vert   \CR \vert \psi_3^I \rangle
\langle \psi_3^I \vert
 \CR  \vert   \psi_1\otimes \psi_2  \rangle
\end{equation}
where the sum runs over an orthonormal basis for
$\CH^{Q_1 + Q_2}_{BPS} $. By varying the final state
$\CF$ one can extract the matrix elements
$\langle \psi_3^I \vert \CR  \vert
\psi_1\otimes \psi_2  \rangle$ and determine
$\CR$, up to unitary rotation in $\CH^{Q_1 + Q_2}_{BPS} $.

 Little is known about the general properties
of the algebra. Examples show that it might
or might not be commutative, associative, Lie,
etc.  It is also worth noting that
the algebra   can depend on the other
kinematic parameters, in particular, on the
``boost direction''  encoded in $t$.

As a simple and explicit example of this
procedure one can use the
the exact $S$-matrices of $d=2, N=2$
integrable systems \cite{fendley}.
We consider the $A_n$ model with
superpotential $W =
{X^{n+1} \over  n+1} -   X$.
The solitons $K_{j,j+r}$ are labelled by $r$ which is an
integer modulo $n$. They have mass
$m_r = {2n \over  n+1} \sin r \mu$, with $\mu=\pi/n$.
BPS multiplets are in small representations of
$N=2$ supersymmetry and therefore
form doublets   $(u_r, d_r)$. The $S$-matrix
is determined by $2 \rightarrow 2$ scattering
and the residue of the $4\times 4$ matrix
for scattering the two-kink state
\begin{equation}
K_{j,j+r} K_{j+r,j+r+s}\rightarrow
 K_{j,j+s}K_{j+s,j+s+r}
\end{equation}
 factors as
$Res = C^\dagger C$ where
$C$ is a certain $2 \times 4$
matrix of rank $2$. For $r+s = 0 \mod n$ the
residue {\it vanishes}.
{}From this one finds  the BPS algebra up
to unitary rotation:
\begin{eqnarray}
&
u_r \cdot u_s   = \sqrt{\sin \mu(r+s)} u_{r+s}
& \nonumber\\
&
d_r \cdot d_s   = 0
& \nonumber\\
&
u_r \cdot d_s   = \sqrt{\sin \mu s } d_{r+s}
& \nonumber\\
&
d_r \cdot u_s   = \sqrt{\sin \mu r} d_{r+s}
& \nonumber\\
\end{eqnarray}
for $r=1,\dots, n-1$ while the product vanishes
for $r+s=0 {\rm mod} n$.
We give this example to illustrate that
the algebra makes sense and is computable
in at least one example. Note
that the algebra structure does not commute
with supersymmetry. (This is a virtue, not
a problem. The BPS algebra might combine
with supersymmetry in an interesting way.)

Two other classes of
examples of infinite-dimensional BPS
algebras are provided by string compactification.
We discuss these in some more detail.

\subsection{Toroidal compactification of heterotic
string}

Dabholkar-Harvey (DH) states are the
perturbative string BPS states of  toroidally
compactified heterotic string \cite{dabholkar}. They
form a subalgebra of the BPS algebra which,
in the tree-level approximation, turns out to
be a GKM \cite{hmii,neumann}.
\footnote{In \cite{Borcherds} R. Borcherds
defined a notion of ``generalized Kac-Moody
algebras.''  The algebra  of DH states is a
further generalization. In order to avoid the
term ``generalized generalized Kac-Moody
algebras'' we reserve the term Borcherds
algebras for the objects defined in
\cite{Borcherds}.}

The residue of a pole in the tree level
$S$-matrix is given by the BRST class
in the operator product of two colliding
on-shell vertex operators. Therefore,
denoting by $\Lambda_\theta$ the action of a
Lorentz boost required to tune to the BPS
pole the product of two (BRST classes of)
vertex operators $V_1, V_2$ is simply:
\begin{equation}
\CR(V_1\otimes V_2)
\equiv
\lim_{z_1 \rightarrow z_2}
\Lambda_{\theta_1}\bigl(V_1(z_1,\bar z_1) \bigr)
\Lambda_{\theta_2}\bigl(V_2(z_2,\bar z_2) \bigr)
\end{equation}
where the product is projected to BRST cohomology.

It is well-known that the operator
product in the BRST cohomology
of a  chiral vertex operator algebra
forms a Lie algebra. (In fact, much more
is true: it is related to ``BV algebras'' and
``Gerstenhaber algebras'' \cite{wittenzwiebach,zuckerman}.)
This is ``just'' the basis of the construction of
gauge symmetry in the heterotic string
\cite{ghmr}.  The above construction also generalizes the
construction of DDF operators.

\subsection{Calabi-Yau algebras}

The BPS states arising in compactifications
of type II string theory on Calabi-Yau
$d$-folds $X$ with $d=2,3,4$
provide another fascinating
class of examples of BPS algebras.

In addition to the perturbative $U(1)$ gauge
bosons arising from Kaluza-Klein reduction
of supergravity fields there are
nonperturbative states
associated to wrapped D-branes.
The  lattice of D-brane charges may
be identified with  the cohomology
lattice of $X$.  The data specifying
a ``wrapped Dbrane state'' on a supersymmetric
cycle $\Sigma \subset X$  includes the data
of a  ``Chan-Paton''
vector bundle on $\Sigma$ with connection.
The Dbrane charges of the state may be
expressed in terms of the characteristic classes
of the Chan-Paton bundle. The BPS
states may then be associated to the
supersymmetric ground states of
a supersymmetric Yang-Mills theory.
In this way
the space of stable
BPS states of RR charge $Q$
can be written in the form:
\begin{equation}
\CH_{BPS}^Q = H^*(\CM(Q)) \otimes \pi
\end{equation}
where $\CM(Q)$ is a certain moduli space
and $\pi$ is a representation of the
supertranslation algebra \cite{polchinski,witten95a,bsv,morrison96,hmii}.
Some aspects of this construction were
anticipated in Kontsevich's work on
mirror symmetry \cite{kontsevich94}.
An attempt was made in
\cite{morrison96,hmii} to give a
precise definition of $\CM(Q)$ but
problems remain. According to
\cite{kontsevich94} one should work
in the bounded derived category.

\subsection{The correspondence conjecture}

{}From the general discussion above
we see that BPS algebras in type II
compactifications must involve some
product structure on the cohomology
spaces $H^*(\CM(Q))$.
An interesting example is provided
by the boundstates of $(0,2,4)$-branes
or of   $(0,2,4,6)$-branes
in compactification of type IIA string on
$K3$ and 3-folds, respectively. In this case
it was argued in \cite{morrison96,hmii} that $\CM(Q)$
is given by a certain moduli space of
sheaves whose characteristic classes
are determined by $Q$. (Essentially,
$Q \sim ch(\iota_{!} E )$ where
$\iota: \Sigma \hookrightarrow X$ is
the inclusion and $E$ is the Chan-Paton
 bundle
on $\Sigma$.
\footnote{This follows from
\cite{ghm} together with some
corrections which will be further discussed in
\cite{minasian}. It suggests, as first
noted by M. Kontsevich and G. Segal,
that conceptually the correct home for
D-brane charges is $K$-theory.})
In this case there is a natural conjecture,
formulated in \cite{hmii}, for the product
structure on the cohomology spaces
called the correspondence conjecture.

We may motivate the correspondence
conjecture as follows. Suppose two
Dbranes collide at a point $P$.
Their respective
Chan-Paton spaces $E_{1} , E_2$ at $P$
must combine to form a new space $E_3$.
However, there are many ways in which this
can be done, each corresponds to an
exact sequence:
$0 \rightarrow E_1 \rightarrow E_3
\rightarrow E_2\rightarrow 0$.  The set of such sequences
is a Grassmannian. When we replace the
Chan-Paton spaces at a point $P$ by Chan-Paton
bundles we obtain a {\it correspondence
subvariety} of
$\CM(Q_1) \times
\CM(Q_2) \times \CM(Q_3)$.
Dbrane groundstates are
 identified with harmonic forms on
$\CM(Q)$ and it is natural to conjecture that the
matrix element $\bigl(\omega_3, \CR(\omega_1 \otimes \omega_2)\bigr) $ is
simply
the overlap integral over the
correspondence variety. Several points in this
proposal must be clarified before it can be taken
to be a precisely formulated conjecture. Some of
these have been addressed in ongoing work with
J. Harvey and D. Morrison. Some aspects of the
enhanced gauge symmetries of $M$-theory
compactification on Calabi-Yau 3-folds are
successfully captured by the correspondence
conjecture \cite{hmm}.

\subsection{Mathematical applications}

The nicest mathematical applications
of BPS algebras are provided by the
heterotic/type II  dual pairs. As we have
seen the heterotic BPS algebras
involve (at least, at tree level) vertex
operator constructions of affine Lie algebras,
Borcherds algebras, and generalizations
thereof.
\footnote{
It is a little mysterious that heterotic tree
level constructions work so well. This might be
related to the pseudo-topological nature of
BPS states. In particular, since their heterotic
vertex operators have right-moving supersymmetric
groundstates one can argue that the BPS product
is not renormalized in perturbation theory. }
 On the other side, the type II
BPS algebras involve products related
to correspondence varieties. These constructions
are very closely related to the work of
Nakajima \cite{Nakajima} (which, of course,
proveded some motivation for the correspondence
conjecture). Indeed, detailed
consideration of subalgebras for
the dual pair of the heterotic string
on $T^4$ and the type IIA string on $K3$
``explains'' Nakajima's geometric constructions
of Heisenberg algebras and affine Lie algebras
\cite{hmii}. The idea that string duality would lead to
a physical explanation of Nakajima's construction
of affine Lie algebras was first published in
\cite{vafa}.

The above philosophy suggests that there are
extensive generalizations of Nakajima's work.
For example, to {\it any} Calabi-Yau threefold
we can associate two infinite dimensional
algebras exchanged by mirror symmetry.
Almost nothing is known about these algebras,
with the exception of the K3-fibered Calabi-Yau's
entering in dual pairs. In this case the subalgebras
associated to boundstates of $(0,2,4)$-branes
in the K3 fiber form  certain infinite-dimensional algebras
which, by string duality, should be related to
vertex operator algebras constructed from
Dabholkar-Harvey
states in $K3 \times T^2$ compactification of
heterotic string. These algebras seem closely
related to the suggestion in \cite{GritNik}
that there are GKM algebras associated to
algebraic K3 surfaces.

A related mathematical application concerns
the relation between automorphic forms and
Calabi-Yau manifolds. These have been
explored in \cite{borcherdsenriques,GritNik,jorgensontodorov}. As we discuss
below,
these forms bear some resemblance to
automorphic forms associated to GKM's.

\subsection{Physical applications}

There are three potential applications of
BPS algebras to physics.

First,
it is an old and haunting idea that the infinite
tower of massive string states form some kind
of infinite tower of generalized massive
gauge bosons for some underlying symmetry
which has been spontaneously broken.
Turning this intuition into precise mathematical
statements has proven rather difficult.
Past attempts include a search for some kind of
``duality invariant string algebra''  or
``universal string symmetry''
\cite{giveon,moore}. It is worth noting that
the BPS algebras associated to Narain
compactifications of heterotic string are
certainly duality invariant, since they are
defined in terms of the physical S-matrix,
while the massless subalgebras are always
the unbroken gauge algebras at all points on
Narain moduli space.  This is one
reason we believe that these algebras will
eventually play a deeper role in the formulation
of string theory.

The dreamy musings of
the previous paragraph are
subject to the sharp criticism that having a
strongly broken gauge symmetry is as useless
as having no symmetry at all. This is correct
in general, however, the systems under study
are very special. (For example, they are often
related to integrable systems.)
 Indeed, as we discuss in the
remainder of the talk, the properties of
BPS states determine the
geometry  of the moduli space of string vacua,
at least for backgrounds with eight
supersymmetries. The connection is made through
the automorphic functions associated
to infinite-dimensional algebras. It is
well-known that the
characters of   affine
Lie algebras define automorphic forms
for subgroups of $SL(2,Z)$. The work of
Borcherds \cite{Borcherds,borcherds95} and of
Gritsenko-Nikulin \cite{GritNik}  shows that
analogous phenomena occur for generalized
Kac-Moody algebras, with $SL(2,Z)$ replaced
by higher rank arithmetic groups.

Third, the automorphic forms related to GKM's
 have also been proposed as counting functions
for counting degeneracies of supersymmetric
black holes \cite{dvvbh}.

\section{Automorphic forms and special geometry }

Automorphic forms of the kind associated to GKM's
appear in the low energy effective actions of
compactified string theories. Particularly
interesting examples are the
 effective couplings
\begin{equation}
\int {1 \over  g^2_{\rm gauge} } \Tr F \wedge * F
\quad {\rm and} \quad \int   {1 \over  g^2_{\rm gravity} }   \Tr R \wedge * R
\end{equation}
in $d=4, N=2$ compactifications of heterotic
and type II strings. These cases are accessible,
yet nontrivial.
The main points which we wish to make about these
couplings are:

First, the quantities are determined entirely by
the BPS states
\cite{hmi,douglasli,bachasfabre,moralesserone,AGNT}.
 Indeed they should be viewed
as regularized sums over the BPS multiplets:
\begin{eqnarray}
&\CF   \sim   \sum_{\rm  \  vm} \CZ^2 \log \CZ^2
- \sum_{\rm  \  hm} \CZ^2 \log \CZ^2
&\nonumber\\
&
{1 \over  g_{\rm gauge}^2}   \sim \sum_{\rm  \  vm} Q_a^2 \log m^2
- \sum_{\rm  \  hm} Q_a^2 \log m^2
&\nonumber\\
&
{1 \over  g_{\rm gravity}^2}   \sim   \sum_{\rm  \  vm}   \log m^2
- \sum_{\rm  \  hm}   \log m^2 &\nonumber\\
\end{eqnarray}
where $\CF$ is the prepotential,
$\CZ$ is the central charge, $Q_a$ is the
gauge charge, and $m$ is the mass.

Second, the resulting formulae
are ideally suited for checking string duality
conjectures. See below.

Third, the quantum corrections
in four dimensions always take the form
(for the nonholomorphic, effective couplings):
\begin{equation}
 {1 \over  g^2(p^2;z)} =
{b \over  16 \pi^2} \log {M_{\rm Planck}^2 \over  p^2} - \log \parallel
\Phi(z)\parallel^2
\end{equation}
where $\Phi(z)$ is an ``interesting'' function
(or section of a line bundle $\CL$ ) over moduli space,
for example, an automorphic form of
$O(2,n;Z)$, $\parallel \cdot \parallel$ is an
invariant norm on $\CL$,  $b$ is the coefficient
of the $\beta$-function, and  $z$ is a point
on moduli space.

\subsection{Threshhold corrections in heterotic
string theory}

Compactification of heterotic string
theory on $K3$ leads to a very interesting
set of automorphic functions. In particular
one can consider compactification on
$K3 \times T^{b_+}$ where $T^{b_+}$
 is a torus of $b_+$ dimensions.
The low-energy theory involves a sigma
model with target space $\CM_{b_+}$
which is the moduli space of vectormultiplets.
The perturbative
corrections to the geometry of
$\CM_{b_+}$ are governed by integrals
of the form \cite{agn,kl,kiritsis,afgnt,dkll,hmi}:
\begin{equation}
\hat F(z) \equiv \int_{\CF} {d^2 \tau \over  y^2 }
 F(q,y) \bar \Theta_{b_+, b_- }(\tau,\bar \tau; z)
\end{equation}
where $\tau$ is the worldsheet modular parameter,
$\CF$ is the fundamental domain
for $PSL(2,Z)$, $y=\Im \tau$,
$F(q,y)$ is a finite sum of
the form
$\sum_{\mu>0} y^{-\mu} F_\mu(q)$ where $F_\mu(q)$
are modular forms.
\footnote{In general for compactifications using
rational conformal field theory one should use
vector valued modular forms. }
$\Theta_{b_+, b_- }(\tau,\bar \tau;z)$ is a
certain sum (depending on the threshold
correction) over
the Narain lattice $\Gamma^{b_+,b_-}$
associated to the flat triples (of metric,
2-form, and Wilson line, $(G,B,A)$) on
$T^{b_+}$, and $z\in  Gr(b_+,b_-)$ is a
point in the Grassmannian determined
by $(G,B,A)$.

\subsection{The $\Theta$-transform}

 Integrals of the type $(13)$
 have been extensively
studied by many authors recently
in both the physics and the  mathematics
literature \cite{hmi,kawai,lust,henningson,borcherds96,kontsevich97}. The
transform
from $F(q,y)$ to $\hat F(z)$ is sometimes
known as a $\Theta$-transform or  ``theta lift.''
The integrals can be done rather
explicitly and the formulae for
$\hat F(z)$  are expressed in terms of a
null-reduced lattice $v^\perp/v$ where
$v\in \Gamma^{b_+,b_-}$ is a null vector.
Physically, a choice of $v$ corresponds to
a choice of decompactification to one higher
dimension.
\footnote{It also corresponds to a choice
of elliptic fibration in the $F$-theory dual.}
Mathematically, it corresponds to
a choice of cusp domain in the Narain moduli
space.

As an example we consider $b_+=2$.
Then there is a well-known  tube domain
realization:
$\CH^{s+1,1}   \equiv  O(s+2,2)/[O(s+2)\times O(2)] \cong
R^{s+1,1}+ i V_+^{s+1,1}$
where $V_+$ is the forward lightcone.
If $F(q)    =\sum_{n=-1}^\infty c(n) q^n
= q^{-1} +c(0) + \cdots$ is a modular form of
weight $-s/2$ then \cite{hmi,borcherds96}:
\begin{eqnarray}
& \widehat{F}(z)
 = - \log \parallel \Phi(z)^2 \parallel^2 &\nonumber\\
& \parallel \Phi(z)\parallel^2 = [-(\Im z)^2] ^{-c(0)/2}
\bigl\vert
\Phi(z) \bigr\vert^2 &\nonumber\\
& \Phi(z) = e^{2 \pi i \rho\cdot z}
\prod_{r>0 }
 \biggl(1-e^{2 \pi i r \cdot z  }
 \biggr)^{c(-r^2/2)}&\nonumber\\
\end{eqnarray}
where the product is (roughly) over $r>0$ in $v^\perp/v$.
Similarly, if we take the transform of
a nonholomorphic form such as $\hat E_2 F(q)$
then the result is an infinite sum of trilogarithms:
\begin{eqnarray}
& \widehat{\hat E_2 F}(z)    = {12 \over  \pi^2 (\Im z)^2} \Biggl\{
 \sum_{r>0}  c(r) \CL_3(e^{2 \pi i r.z})+ \cdots \Biggr\} &\nonumber\\
& \cdots + {1 \over  (Im z)^2} \bigl[\tilde d_{ABC}(\Im z)^A (\Im z)^B (\Im
z)^C + {6\zeta(3) \over  \pi^2} c(0) \bigr]
&\nonumber\\
\end{eqnarray}
here $\CL_3$ is the ``Bloch-Ramakrishnan-Wigner''
polylogarithm:
\footnote{Bloch and Wigner refer to the mathematicians,
not the physicists of that name!}
\begin{equation}
\CL_3(z) = \Re\Biggl[Li_3(z) - \log\vert z \vert Li_2(z)
+ {1 \over  3} \log^2\vert z \vert Li_1(z)
\Biggr]
\end{equation}
Moreover, these integrals behave nicely under
the action of invariant differential operators
on $\CH^{s+1,1}$. For example, we have
\begin{equation}
(\Delta_z - (s+4) ) \widehat{\hat E_2 F} =  6 (s+4)
\log \parallel \Psi(z)  \parallel^2 +const.
\end{equation}
where $\Delta_z$ is the invariant Laplacian
and $\Psi(z)$ is an automorphic product.

The method of evaluation of $\hat F(z)$ is
straightforward, although the details are a little
intricate. Choosing a null vector one uses
the Poisson summation formula to convert
a subsum in $\Theta$
 over vectors in a hyperbolic lattice
$\Gamma^{1,1}$ to a sum over a pair of integers
$(w_1,w_2)$ which - in the language of conformal
field theory of periodic bosons on the worldsheet
torus - may be regarded as winding numbers
around the $a,b$-cycles. One then separates the
sum over $(w_1,w_2)$ into a sum over
$\ell = g.c.d.(w_1,w_2)$ and a sum over relatively
prime integers. The latter sum can be used to
unfold the fundamental domain to the strip. The
sum on $\ell$ then becomes an infinite sum of
Bessel functions.
\footnote{In the example of gauge coupling
threshhold corrections these Bessel functions
are Green's functions, as expected once we
identify $\Im \tau$ with a Schwinger parameter.
In four-dimensional case of
$b_+=2$ the sum produces the logarithmic
series $\log[1-e^{2\pi i r\cdot z}]$. Thus, the
logarithmic couplings and trilogarithmic
prepotentials arise from sums over the
Kaluza-Klein towers. This explains -
at least technically - the relation to the trilogarithms
which occur in N. Nekrasov's work on five dimensional
super Yang-Mills theory \cite{nekrasov}. }

Part of the interest of the
integrals $(13)$  is that they are, manifestly,
invariant under the arithmetic
group $O(\Gamma^{b_+,b_-})$. We can
thus learn things about automorphic forms.
For example, for $b_+=2$ we see that the
integral is almost holomorphically split.
The nonholomorphic term $\sim \log (\Im z)^2$
shows that $\Phi(z)$ is in fact an automorphic
form of weight $c(0)/2$. The zeroes and poles
of $\Phi(z)$ are easily found since the
singularities of $\hat F(z)$ always arise
from the divisors
where BPS states become massless. In this
way we can reproduce the results of
\cite{borcherds95}.

An important point is that the integrals show
{\it chamber dependence}. That is $\hat F$
(or some derivative) is discontinuous across
 real codimension one subvarieties of
$Gr(b_+,b_-)$. This is the technical source of
chamber dependence of prepotentials and
gauge couplings in 4 and 5 dimensional
heterotic string compactifications.

\subsection{Results for heterotic compactifications}

In the case of four-dimensional compactifications
one defines a   vector valued modular form from
the trace in the internal superconformal field theory:
\begin{equation}
{i \over  2} {1 \over  \eta^2} \Tr_R J_0 e^{i \pi J_0}
q^{H}
\bar q^{\tilde H}
= \sum_i Z^i_\Gamma(q,\bar q) f_i(q)
\end{equation}
where $ f_i(q) = \sum c_i(x) q^x$ are modular  forms for $\Gamma(N) \subset
SL(2,Z)$ and
$Z^i_\Gamma(q,\bar q) \sim \sum q^{\half p_L^2}
\bar q^{\half p_R^2} $ are sums over cosets of the
Narain lattice. Following the
one-loop analysis of
\cite{agn,kl,kiritsis,afgnt,dkll,hmi,kawai,lust,henningson}
one finds the perturbative
prepotential is given by:
\begin{eqnarray}
& \CF^{\rm Het} (\tau_S,z)   =\half  \tau_S z^2  +
 { 1\over  3!} d_{ABC} z^A z^B z^C
+ {i \zeta(3)   \over
 (2\pi)^3} \Delta n  & \nonumber\\
& +  { 1 \over  (2 \pi i )^3 }
  \sum_{r>0,i} c_i(-r^2/2 )  Li_3( e^{    2 \pi  i  r\cdot z } )
+ \CO( e^{ 2 \pi i \tau_S} )& \nonumber\\
\end{eqnarray}
where $\tau_S$ is the heterotic axiodil,
$\Delta n$ is the net number of
massless vector-multiplets minus
hyper-multiplets,
$z$ is a point in the tube domain,
the sum on $r>0$ is (roughly) a sum on vectors
in $v^\perp/v$ in the forward light cone,
and $d_{ABC}$ is a chamber-dependent
symmetric tensor.
Using  this expression and
identities such as $(17)$ one finds
expressions such as $(12)$ for the
gauge couplings. Similar results hold
for gravitational couplings.

In the case of five-dimensional compactifications
($b_+=1$) one derives piecewise linear
gauge couplings and piecewise cubic
prepotentials similar to  those that have
been studied in \cite{AFT,ms,dkv,ims}.
The connection between the 4 and 5-dimensional
results can be obtained by taking an appropriate
limit of the moduli. In terms of the standard
moduli $T=  2 i R_2/ R_1$,$U=  i R_2 R_1$
for $T^2$
in $K3 \times T^2$ compactification one finds
that the prepotential and coupling behave
as: $\CF \rightarrow {1 \over  3!} \tilde d_{ABC} z^A z^B z^C$
and ${1 \over  g^2} \rightarrow (\tilde \rho, z) $
where $\tilde \rho$ is obtained from
the Weyl vector for the
infinite product occuring in $(15)$.

A very curious point is that the $\Theta$-transform
turns out to be closely related to Donaldson
invariants \cite{moorewitten},
as formulated in Donaldson-Witten
theory \cite{witten88,witten94}.
(Such a connection had been guessed in
\cite{borcherds96}.)
In particular, the linear gauge couplings
studied in \cite{ms,dkv,ims} correspond to
one-point functions of a 2-observable in
the theory of a Euclidean 3-brane wrapping
a Del-Pezzo base of an elliptically fibered
Calabi-Yau. Unfortunately, it is not clear
{\it a priori} why there should be a relation
between these two quantities.

\subsection{Applications to duality}

Equation $(19)$ should be compared to the
famous nonperturbatively exact
expression (almost equivalent to the multiple
cover formula)  for the prepotential for
special Kahler geometry of type II compactification
\begin{eqnarray}
&
\CF^{\rm Type II}   =
{ N_{ABC} \over  3!}  z^A z^B z^C  + {i \zeta(3) \chi \over
2 (2\pi)^3} &\nonumber\\
&+
{1 \over  (2 \pi i)^3}
\sum_{r>0} n(r)   Li_3( e^{2 \pi i r \cdot z} )
&\nonumber\\
\end{eqnarray}
where $z$ are flat coordinates in the
complexified K\"ahler cone, the
sum $r>0$ is over the subcone of the
Mori cone generated by the
rational curves, and $N_{ABC}$ are classical
intersection numbers.

Comparison with the previous heterotic
results gives concrete and nontrivial evidence
for string dual pairs such as those
proposed in \cite{kachruvafa}: the vector multiplet $\tau_S$
is identified with the volume of the base of the
IIA K3 fibration \cite{klm,aspinwalllouis}.
The nontrivial fact that the
heterotic   expression, valid to all orders of
perturbation theory in $e^{2 \pi i \tau_S}$,
matches a IIA expression allows one to define
the nonperturbative heterotic prepotential to
be the IIA expression.  The  approach
of direct calculation of the heterotic prepotential
complements the approaches in
\cite{klm,klt,afgnt95b}.

We can turn the above reasoning  around.
Assuming
string duality one can use heterotic computations
to make very nontrivial predictions about the
geometry of $K3$ surfaces.
For example, the heterotic dual of a simple
$Z_2$ orbifold studied in \cite{hmi} involves
the Calabi-Yau manifold
$x_1^{84} + x_2^{84} + x_3^7 + x_4^3 + x_5^2 =0$
in $P_4^{1,1,12,28,42}$
with K3 fibers $x_1^{42} +  x_3^7 + x_4^3 + x_5^2 =0 $
in $P_3^{1,6,14,21}$.
String duality predicts that the holomorphic rational
curves in such K3 surfaces should be ``counted''
\footnote{In an appropriate sense, which has
yet to be fully elucidated. The curves will
come in families and the counting of the
family probably involves
an integral of an Euler class of an ``antighost'' bundle. }
by the modular form
$E_6/\eta^{24}$. This should be contrasted with
the formula $1/\eta^{24}$ which counts rational
curves in $K3$, which are holomorphic in {\it some}
complex structure compatible with a fixed
hyperkahler structure
\cite{bsv,yauzaslow,beauville}.

As another example of a (new) result along these lines
one can combine the techniques of \cite{borcherds96}
with the integral formula of
\cite{AGNT} to calculate the one-loop heterotic
formula for the threshhold correction $F_g$.
The result is an expression of the form:
\begin{eqnarray}
& F_g =  \Re\Biggl[
\sum_{r>0} \sum_{t=0}^g
c(r^2/2, t) & \nonumber\\
&\biggl\{ \CP_0
\log[1-e^{2 \pi i r\hat\cdot z} ] +
\CP_1 {1 \over  1-e^{2 \pi i r\hat\cdot z} }
& \nonumber\\
&+ \cdots
 + \CP_{2g-2} \biggl(
{1 \over  1-e^{2 \pi i r\hat\cdot z}}\biggr)^{2g-2} \biggr\}
\Biggr]
& \nonumber\\
\end{eqnarray}
Here  $c(r^2/2, t)$ are coefficients of certain
modular forms of weight $g$ which have the
form:
\begin{eqnarray}
{E_4 E_6 \over \eta^{24}}\Biggl(
{1 \over  g!} ({\pi^2\over  3}  \hat E_2)^g  + \cdots
+ {2 \zeta(2g) \over  g} E_{2g}\Biggr)
& = \nonumber\\
 \sum c(m,t) q^m y^{-t} &
& \nonumber\\
\end{eqnarray}
($E_{2k}(\tau)$ are Eisenstein series of weight $2k$),
  $\CP$ are Laurent polynomials in $(\Im z)^2, \Im(r\cdot z)$
and $r \hat{\cdot} z \equiv \Re(r \cdot z) + i \vert \Im (r \cdot z) \vert$

According to \cite{bcov} a certain holomorphic
function of $z$ extracted from equation
$(21)$
will compute the number of holomorphic
genus $g$ curves in the K3 fiber of the
Calabi-Yau in the dual type IIA compactification.

\section{Statecounting for black holes}

In \cite{dvvbh}  R. Dijkgraaf, E. Verlinde, and
H. Verlinde proposed a counting formula for
the
degeneracies of black holes preserving
$1/4$ of the supersymmetry of the background
 IIA/$K3 \times T^2$, or, equivalently,
heterotic/$T^6$. Their formula involves
an automorphic form of a GKM in an
intriguing way.

Specifically, using a computation by
T. Kawai \cite{kawai95} of a particularly interesting $\Theta$-transform
which converts the K3 elliptic genus
$\chi_{\tau,z}(K3) = \sum_{h\geq 0,\, m\in Z}
c(4h-m^2)e^{2\pi i (h\tau+mz)}
$
to the automorphic form \cite{GritNik}:
\begin{eqnarray}
&\Phi(T,U,V)= e^{2\pi i(T +U +V)} \cdot &\nonumber\\
& \cdot \prod_{(k,l,m)> 0}\left(1 -e^{2\pi
i(kT+l U+m V)}\right)^{c(4kl-m^2)},  &\nonumber\\
\end{eqnarray}
which is the denominator product for the
GKM based on the Cartan matrix:
\begin{equation}
A= \pmatrix{2 & -2 &-2 \cr
-2 & 2 & -2 \cr
-2 & -2 & 2 }\quad ,
\end{equation}
\cite{dvvbh}
 proposed a counting formula for
the supersymmetric black holes with
(electric, magnetic) charge vectors
$(q_e, q_m) \in II^{6,22} \oplus II^{6,22}$:
\begin{equation}
d(q_e, q_m) =  \oint_0^1 d T d U d V \,
{e^{i \pi (q_m^2  T + q_e^2 U +
2q_e\cdot q_m V)} \over \Phi( T, U, V)}
\end{equation}
Their argument
used ideas about six-dimensional
strings theories.

Quite generally the elliptic genera of symmetric products
of K\"ahler manifolds $S^N X$ are summarized by an
infinite product formula \cite{dvvbh,dmvv}:
\footnote{generalizing the famous
result of \cite{gottsche,hirzebruch,vafawitten}}
\begin{eqnarray}
& \sum_{N=0}^\infty p^N \chi(S^N X;q,y) &\nonumber\\
& = \prod_{n>0,m\geq 0,\ell}
{1\over (1-p^n q^m y^\ell)}{}_{c(nm,\ell)} &\nonumber\\
\end{eqnarray}
where the coefficients are defined by
the elliptic genus of $X$:
$\chi( X; q,y) = \sum_{m\geq 0,\ell} c(m,\ell) q^m y^\ell$.
In view of the work of Strominger and Vafa on
black hole entropy \cite{sv} we may expect
black hole degeneracies to be counted by
similar denominator products in more general
situations.

\section{Conclusion: Search for the missing synthesis}

The reader will, no doubt, find this talk
unsatisfying because, while sections 2,3,
and 4 are clearly closely related, we have
given no precise explanation of {\it how} they
are related.  Indeed, the results in sections
3 and 4 can be, and often are, presented without
any mention of BPS algebras at all.
Nevertheless, while BPS algebras have
so far played no significant role in string
theory we believe they have the potential
to become a useful, and even important
tool. In order to justify this belief we outline
a speculation \cite{hmi} based on the relation between
the prepotentials $(19)$ and $(20)$  to the prepotentials
of Seiberg-Witten theory \cite{seibergwitten}.

Consider the theory of a
$d=4, N=2$ supersymmetric
vectormultiplet  for a compact
gauge group $G$ with Lie algebra
$\lieg$. In the semiclassical
regime the Coulomb branch
is $\liet \otimes C/W$ where
$\liet$ is the Cartan subalgebra and
$W$ is the Weyl group. The
perturbative prepotential is:
\begin{equation}
\CF^{\rm rigid,1-loop}
 = - { 1 \over  8\pi^2}
\sum_{\vec \alpha>0}
(\vec \alpha\cdot \vec a)^2
\log{(\vec \alpha\cdot \vec a)^2 \over  \Lambda^2 }
\end{equation}
where the sum is taken over the positive
roots of $\lieg$.
Let us compare this with the perturbative
prepotential $(19)$. Letting $M_{st}$ denote the
string mass we can take the limit from local
to rigid special K\"ahler geometry
by taking:
$z  = z_0 + (  {\vec a \over  M_{\rm st }}; 0,0)$
and letting $M_{\rm st } \rightarrow \infty$ at
fixed $\vec a$, $z_0 = (\vec 0; T,U)$.  Using
\begin{equation}
Li_3 (1-x)  \rightarrow -{1\over  2} x^2 \log x + \CO(x^3 \log x)
\end{equation}
and suitably renormalizing $\tau_S$
we recover the expression
\begin{equation}
i \CF  = \tilde \tau_S \vec a^2
- {1 \over  8 \pi^2 } \sum_{\vec \alpha >0}
(\vec \alpha \cdot \vec a)^2 \log\bigl[
{2 \pi i \vec \alpha \cdot \vec a
\over  M_{\rm st } } \bigr]
\end{equation}
We thus learn that the trilogarithms in
$(19)$ should be
viewed as stringy deformations of the logarithms
of Seiberg-Witten theory \cite{hmi}.
Moreover,  the prepotential  $(20)$,  which is the
nonperturbative completion of $(19)$,  is exact.
\footnote{In contrast to string theory,
the exact Seiberg-Witten prepotential
does not seem to be usefully expressed in
terms of a sum of logarithmic functions. For a
related context where this is true see
\cite{nekrasov}.}
This raises the idea \cite{hmi}
that the nonperturbative
quantum corrections are completely determined
by some algebraic structure, presumably,
something like the algebra of BPS states.
The sum over rational curves in  $(20)$ should
be viewed as a sum over positive roots.
The numbers of rational curves are root
degeneracies. The analog of the complexified
Cartan algebra is the complexified Kahler cone.
The gauge couplings $\tau_{IJ}$ are, quite
generally, logarithms of ``automorphic products''
constructed using the geometry of the
Calabi-Yau 3-fold and generalizing the
denominator products of GKM's.
Speculating further an
algebraic foundation for determining
the geometry of string moduli spaces could
lead to classification of string vacua, perhaps
along the lines of the finiteness results of
\cite{nikulin}.

We find the above ideas and speculations
attractive,  but they clearly suffer from the
imprecision that no prescription is given for
telling how to sum over BPS states, and with
what weighting. With this in mind
with J. Harvey
we investigated in \cite{hm96b,hm96c}
certain gravitational threshold corrections
which seem to be particularly closely related
to known GKM's. Unfortunately, the precise
relation of these corrections to the BPS
algebra remains unclear, although the
result of \cite{hm96c} is quite suggestive.
This paper shows that the gravitational
correction $F_1$ for the
dual pair of \cite{fhsv} is determined by
the denominator product of a certain
generalized Kac-Moody superalgebra
studied in \cite{borcherdsenriques}.

In conclusion, the work we have reviewed has
lead to some interesting new results on quantum
corrections and automorphic forms. It might lead
to important geometrical realizations of
affine Lie algebras and generalized Kac-Moody
algebras. However,
our main goal: showing
 that new algebraic structures,
generalizing Kac-Moody algebras,
are fundamental to string duality, remains
elusive. This idea has the potential to become
extremely useful, but at present there are only
hints to support its validity.

\section{Acknowledgements}

We would like to thank J. Harvey for a
very valuable and stimulating
collaboration that lead to
the above results and ideas.
We would also  like to
thank R. Borcherds, R. Dijkgraaf,
A. Gerasimov,  S. Katz,  A. Losev,
R. Minasian, D. Morrison, N. Nekrasov,
S. Shatashvili,
A. Todorov, E. Verlinde, H. Verlinde,
E. Witten, and G. Zuckerman for very helpful
discussions on the above topics .


\begin{thebibliography}{9}
\bibitem{agn}{I. Antoniadis,  E. Gava, K.S. Narain, ``Moduli corrections to
gravitational
couplings from string loops,'' Phys. Lett. {\bf B283} (1992) 209,
hep-th/9203071; `` Moduli corrections to gauge and gravitational couplings in
four-dimensional superstrings,'' Nucl. Phys. {\bf B383} (1992) 109,
hep-th/9204030.}
\bibitem{afgnt}{I. Antoniadis, S. Ferrara, E. Gava, K.S. Narain and T.R.
Taylor,
``Perturbative Prepotential and Monodromies in N=2 Heterotic Superstring,''
Nucl. Phys. {\bf B447} (1995) 35, hep-th/9504034.}
\bibitem{afgnt95b}{I. Antoniadis, S. Ferrara, E. Gava, K.S. Narain and T.R.
Taylor,``Duality Symmetries in N=2 Heterotic Superstring,''
hep-th/9510079;Nucl.Phys.Proc.Suppl. 45BC (1996) 177-187; Nucl.Phys.Proc.Suppl.
46 (1996) 162-172}
\bibitem{AFT}{I. Antoniadis, S. Ferrara, T.R. Taylor,
``N=2 Heterotic Superstring and its Dual Theory in Five Dimensions,''
hep-th/9511108; Nucl.Phys. B460 (1996) 489-505.}
\bibitem{AGNT}{I. Antoniadis, E. Gava,
K. S. Narain and T. R. Taylor, ``$N=2$ Type
II- Heterotic duality and higher derivative F-terms, '' hep-th/9507115;  M.
Serone,
``N=2 Type I-Heterotic Duality and Higher Derivative F-Terms,''
hep-th/9611017; hep-th/9611017}
\bibitem{aspinwalllouis}{P. Aspinwall and J. Louis,
``On the Ubiquity of K3 fibrations
in string duality,'' Phys. Lett. {\bf B369} (1996) 233; hep-th/9510234.}
\bibitem{bachasfabre}{C. Bachas and C. Fabre,
``Threshold effects in open-string theory,'' hep-th/9605028;
Nucl.Phys. B476 (1996) 418-436}
\bibitem{beauville}{A. Beauville, ``Counting rational curves on K3 surfaces,''
alg-geom/9701019}
\bibitem{bcov}{M. Bershadsky, S. Cecotti, H. Ooguri and C. Vafa, ``
Kodaira-Spencer
theory
of gravity and exact results for quantum string amplitudes, '' Commun. Math.
Phys.
{\bf 165} (1994) 311, hep-th/9309140. }
\bibitem{bsv}{M. Bershadsky, V. Sadov, and
C. Vafa, ``D-Strings on D-Manifolds,''  hep-th/9510225;
Nucl.Phys. B463 (1996) 398-414;  C. Vafa,
``Gas of D-Branes and Hagedorn Density of BPS States,''
hep-th/9511088;Nucl.Phys. B463 (1996) 415-419;
M. Bershadsky, V. Sadov, and
C. Vafa, ``D-Branes and Topological Field
Theories,'' Nucl. Phys. {\bf B463} (1996) 420; hep-th/9511222.}
\bibitem{Borcherds}{R. E. Borcherds,  ``Generalized Kac-Moody algebras,''
Journal of
Algebra {\bf 115} (1988) 501;
``The monster Lie algebra,'' Adv. Math. {\bf 83}
No. 1 (1990);``Monstrous moonshine
and monstrous Lie superalgebras,'' Invent. Math.
{\bf 109}(1992) 405.}
\bibitem{borcherds95}{R. E. Borcherds, ``Automorphic forms
on $O_{s+2,2}(R)$ and infinite products,''
Invent. Math. {\bf 120}(1995) 161;`Automorphic forms
on $O_{s+2,2}(R)^+$ and generalized Kac-Moody
algebras,''  Proc. of the 1994 International
 Congress of Mathematicians,
p.744, Birkh\"auser, 1995}
\bibitem{borcherdsenriques}{R. E. Borcherds, ``The moduli space
of Enriques surfaces and the fake monster Lie
superalgebra,'' Topology vol. 35 no. 3, (1996) 699.}
\bibitem{borcherds96}{R.  E. Borcherds,  ``Automorphic forms with singularities
on
Grassmannians,'' alg-geom/9609022}
\bibitem{lust}{G. L. Cardoso ,  G. Curio ,  D. Lust ,  T.
Mohaupt ,  S.-J. Rey, ``BPS Spectra and Non--Perturbative Couplings in N=2,4
Supersymmetric String Theories,''  hep-th/9512129;
G. L. Cardoso ,  G. Curio ,  D. Lust ,  T.
Mohaupt ,
``Instanton Numbers and Exchange Symmetries in $N=2$ Dual String Pairs,''
hep-th/9603108; G. L. Cardoso ,  G. Curio ,
D. Lust ,
``Perturbative Couplings and Modular Forms in N=2 String Models with a Wilson
Line,'' hep-th/9608154}
\bibitem{dabholkar}{A. Dabholkar and J. Harvey,
``Nonrenormalization of the superstring tension,''
Phys.Rev.Lett.63:478,1989.}
\bibitem{dkll}{B. de Wit, V. Kaplunovsky, J. Louis and D.  L\"{u}st,
``Perturbative Couplings of Vector Multiplets in $N=2$ Heterotic String
Vacua,'' Nucl. Phys. {\bf B451} (1995) 53, hep-th/9504006.}
\bibitem{dvvbh}{R. Dijkgraaf,
E. Verlinde, and H. Verlinde, ``Counting
dyons in N=4 string theory,'' hep-th/9607026;
Nucl.Phys. B484 (1997) 543-561}
\bibitem{dmvv}{R. Dijkgraaf, G, Moore,
E. Verlinde, and H. Verlinde,
``Elliptic Genera of Symmetric Products and Second Quantized Strings,''
hep-th/9608096; Commun.Math.Phys. 185 (1997) 197-209}
\bibitem{dkv}{M. Douglas, S. Katz, and C. Vafa,
``Small Instantons, del Pezzo Surfaces and Type I' theory,''
hep-th/9609071; Nucl.Phys. B497 (1997) 155-172}
\bibitem{douglasli}{M. R. Douglas  and  M. Li,
``D-Brane Realization of N=2 Super Yang-Mills Theory in Four Dimensions,''
hep-th/9604041}
\bibitem{fendley}{P. Fendley and K. Intriligator,  ``Scattering
and thermodynamics in integrable $N=2$ theories,''
hep-th/9202011}
\bibitem{fhsv}{S. Ferrara, J. A. Harvey, A. Strominger, C. Vafa ,
``Second-Quantized Mirror Symmetry, '' Phys. Lett. {\bf B361} (1995) 59;
hep-th/9505162. }
\bibitem{giveon}{A. Giveon and M. Porrati, ``Duality invariant string algebra
and
$D=4$ effective actions, '' Nucl. Phys. {\bf B355} (1991) 422.}
\bibitem{gottsche}{L. G\"ottsche,
Lecture Notes in Mathematics 1572,
Hilbert Schemes of Zero-Dimensional Subschemes of Smooth Varieties,
Springer-Verlag, Berlin 1994.}
\bibitem{ghm}{M.B. Green,  J.H. Harvey, and
G. Moore, ``I-Brane inflow and Chern-Simons couplings
on D-branes,'' hep-th/9605033;Class.Quant.Grav. 14 (1997) 47-52}
\bibitem{GritNik}{V. A. Gritsenko, V. V. Nikulin,
``Siegel automorphic form corrections of some Lorentzian Kac--Moody Lie
algebras, '' alg-geom/9504006;
V. A. Gritsenko, V. V. Nikulin, ``K3 Surfaces,
Lorentzian Kac-Moody Algebras, and
Mirror Symmetry,'' alg-geom/9510008;
``Automorphic Forms and Lorentzian Kac--Moody Algebras. Part I,''
alg-geom/9610022;
``Automorphic Forms and Lorentzian Kac-Moody Algebras. Part II,''
alg-geom/9611028;
``A lecture on Arithmetic Mirror Symmetry and Calabi-Yau manifolds,''
alg-geom/9612002.
}
\bibitem{ghmr}{D. Gross, J. Harvey, E.  Martinec, and
R. Rohm, ``Heterotic string theory I. The free heterotic string,''
Nucl.Phys.B256:253,1985; ``Heterotic string theory II. The interacting
 heterotic string,'' Nucl.Phys.B267:75,1986.
}
\bibitem{hmi}{J. Harvey and G. Moore, ``Algebras, BPS States, and Strings,''
hep-th/9510182;
Nucl.Phys. B463 (1996) 315-368}
\bibitem{hmii}{J. Harvey and G. Moore, ``On the algebra of BPS states,''
hep-th/9609017}
\bibitem{hm96b}{J. Harvey and G. Moore,
``Fivebrane Instantons and $R^2$ couplings in $N=4$ String Theory,''
hep-th/9610237}
\bibitem{hm96c}{J. Harvey and G. Moore, ``Exact Gravitational Threshold
Correction in the FHSV Model,'' hep-th/9611176}
\bibitem{hmm}{J. Harvey, G. Moore, and D. Morrison, in progress.}
\bibitem{henningson}{M. Henningson and G. Moore,
`` Counting Curves with Modular Forms,'' hep-th/9602154;
``Threshold corrections in $K3\times T2$
heterotic string compactifications.'' hep-th/9608145.}
\bibitem{hirzebruch}{F. Hirzebruch and T. Hofer, Math. Ann. 286 (1990)255.}
\bibitem{ims}{K. Intriligator, D. Morrison, and
N. Seiberg,
``Five-Dimensional Supersymmetric Gauge Theories and Degenerations of
Calabi-Yau Spaces,'' hep-th/9702198; Nucl.Phys. B497 (1997) 56-100}
\bibitem{jorgensontodorov}{J. Jorgenson
and A. Todorov, ``A conjectured analog of Dedekind's eta function for K3
surfaces,'' Yale preprint;
``Enriques surfaces, analytic
discriminants, and Borcherd's $\Phi$ function,'' Yale preprint;
}
\bibitem{kachruvafa}{S. Kachru and C. Vafa, ``Exact results for $N=2$
compactifications of
heterotic strings, '' Nucl. Phys. {\bf B450} (1995) 69; hep-th/9505105.}
\bibitem{kl}{V. Kaplunovsky, ``One loop threshold effects in
string unification,'' hep-th/9205070,
Nucl. Phys. {\bf B307} (1988) 145;
L. Dixon,  V. S. Kaplunovsky and J. Louis, ``Moduli-dependence of string
loop corrections to gauge coupling constants, ''Nucl. Phys. {\bf B329} (1990)
27;V. Kaplunovsky and J. Louis, ``On gauge couplings in string
theory,''
Nucl. Phys. {\bf B444} (1995) 191, hep-th/9502077}
\bibitem{klt}{V. Kaplunovsky, J. Louis and S. Theisen, ``Aspects of duality in
$N=2$ string vacua,'' Phys. Lett. {\bf B357} (1995) 71, hep-th/9506110.}
\bibitem{kawai95}{T. Kawai,
 ``$N=2$ heterotic string threshold correction,
$K3$ surface and generalized Kac-Moody superalgebra,'' hep-th/9512046; C.D.D.
Neumann,
``The elliptic genus of Calabi-Yau 3- and 4-folds, product formulae and
generalized Kac-Moody algebras,'' hep-th/9607029}
\bibitem{kawai}{T. Kawai, ``String
Duality and Modular Forms,'' hep-th/9607078.}
\bibitem{kiritsis}{E. Kiritsis and C. Kounnas,
``One Loop Corrections to Coupling Constants in IR-regulated String Theory,''
hep-th/9509017,Nucl.Phys.Proc.Suppl. 45BC (1996) 207-216;
E. Kiritsis, C. Kounnas, M. Petropoulos, J. Rizos,
``On the Heterotic Effective Action at One-Loop, Gauge Couplings and the
Gravitational Sector,'' hep-th/9605011;``Universality properties of N=2 and N=1
Heterotic threshold corrections,'' hep-th/9608034, Nucl.Phys. B483 (1997)
141-171.}
\bibitem{klm}{A. Klemm, W. Lerche and P. Mayr, ``K3-fibrations and
Heterotic-Type II string duality, '' Phys. Lett. {\bf B357}
(1995) 313, hep-th/9506122.}
\bibitem{kontsevich94}{M. Kontsevich,
``Homological Algebra of Mirror
Symmetry,'' Proc. of the 1994 International
 Congress of Mathematicians,
p.120, Birkh\"auser, 1995; alg-geom/9411018}
\bibitem{kontsevich97}{M. Kontsevich,
``Product formulas for modular forms on
$O(2,n)$,'' alg-geom/9709006, Sem. Bourbaki
1996, no. 821}
\bibitem{zuckerman}{B. H. Lian and  G. J. Zuckerman, ``New
Perspectives on the BRST Algebraic Structure of
String Theory,''  Commun.Math.Phys. {\bf 154} (1993) 613;
hep-th/9211072.}
\bibitem{minasian}{R. Minasian and G. Moore, to appear}
\bibitem{moore}{G. Moore,
``Finite in All Directions, '' hep-th/9305139;  ``Symmetries and
symmetry-breaking in string theory,'' hep-th/9308052;
``Symmetries of the Bosonic String S-Matrix,''
hep-th/9310026; Addendum to: ``Symmetries of the Bosonic String S-Matrix,''
hep-th/9404025.}
\bibitem{moorewitten}{G. Moore and E. Witten,
``Integration over the $u$-plane in Donaldson
theory,'' hep-th/9709193.}
\bibitem{moralesserone}{J.F. Morales and M. Serone,
``BPS states and supersymmetric index in N=2 type I string vacua,''
hep-th/9703049.}
\bibitem{morrison96}{D. Morrison,
``The Geometry Underlying Mirror Symmetry,''
alg-geom/9608006;
Proc. European Algebraic Geometry Conference (Warwick, 1996).}
\bibitem{ms}{D. Morrison and N. Seiberg,
``Extremal Transitions and Five-Dimensional Supersymmetric Field Theories,''
hep-th/9609070; Nucl.Phys. B483 (1997) 229-247}
\bibitem{Nakajima}{H. Nakajima, ``Instantons on ALE spaces,
quiver varieties, and Kac-Moody algebras,'' Duke Math.
{\bf 76} (1994)365;
``Gauge theory on resolutions of simple singularities
and simple Lie algebras,'' Intl. Math. Res. Not.
{\bf 2}(1994) 61;
``Quiver Varieties and Kac-Moody algebras,''
preprint; ``Heisenberg algebra and Hilbert
schemes of points on projective surfaces,''
alg-geom/9507012; ``Instantons and
affine Lie algebras,'' alg-geom/9502013;
``Lectures on Hilbert schemes of points on
surfaces,'' preprint.}
\bibitem{nekrasov}{N. Nekrasov, ``Five Dimensional Gauge Theories and
Relativistic Integrable Systems,'' hep-th/9609219.}
\bibitem{neumann}{C.D.D. Neumann,
``Perturbative BPS-algebras in superstring theory,'' hep-th/9702197;
Nucl.Phys. B499 (1997) 596-620}
\bibitem{nikulin}{V. V. Nikulin,
``Reflection groups in hyperbolic spaces and the
denominator formula for Lorentzian Kac--Moody Lie algebras,''
alg-geom/9503003}
\bibitem{polchinski}{J. Polchinski, ``Dirichlet-Branes and Ramond-Ramond
charges''
hep-th/9510017; S. Chaudhuri, C.Johnson and J. Polchinski, ``Notes on
D-branes,'' hep-th/9602052; J. Polchinski,
``TASI Lectures on D-Branes,'' hep-th/9611050.}
\bibitem{seibergwitten}{N. Seiberg and E. Witten,
``Monopole Condensation, And Confinement In $N=2$ Supersymmetric Yang-Mills
Theory,''
hep-th/9407087;Nucl. Phys. {\bf B426} (1994) 19;
`Monopoles, Duality and Chiral Symmetry Breaking in N=2 Supersymmetric QCD,''
hep-th/9408099;Nucl. Phys. {\bf B431} (1994) 484.}
\bibitem{sv}{A. Strominger and   C. Vafa, ``Microscopic Origin of the
Bekenstein-Hawking Entropy,'' hep-th/9601029; Phys.Lett. B379 (1996) 99-104}
\bibitem{vafa}{C. Vafa, ``Instantons on D-branes,'' Nucl. Phys. {\bf B463}
(1996)
435; hep-th/9512078.}
\bibitem{vafawitten}{C. Vafa and E. Witten,
``A strong coupling test of S-duality,''
hep-th/9408074; Nucl. Phys. {\bf B431}(1994)3-77}
\bibitem{witten88}{E. Witten,
``Topological Quantum Field Theory,''
Commun. Math. Phys. {\bf 117} (1988)
353.}
\bibitem{wittenzwiebach}{E. Witten and
B. Zwiebach, ``Algebraic Structures and Differential Geometry in 2D String
Theory,''
hep-th/9201056, Nucl. Phys. B377 (1992) 55-112}
\bibitem{witten94}{E. Witten, ``Monopoles and
four-manifolds,'' hep-th/9411102,
 Math. Research Letters {\bf 1} (1994) 769.}
\bibitem{witten95a}{E. Witten, ``Bound States Of Strings And $p$-Branes,''
hep-th/9510135; Nucl.Phys. B460 (1996) 335-350}
\bibitem{yauzaslow}{S.T. Yau and E. Zaslow,
``BPS States, String Duality, and Nodal Curves on K3,'' hep-th/9512121;
Nucl.Phys. B471 (1996) 503-512}
\end{thebibliography}
\end{document}